\documentclass[preprint,12pt]{elsarticle}

\usepackage{amsmath,amssymb,amsfonts}

\usepackage{graphicx}
\usepackage{xcolor}

\usepackage{array}

\usepackage{xurl}
\usepackage{hyperref}
\hypersetup{breaklinks=true}

\usepackage{tikz}
\usetikzlibrary{shapes,arrows}

\journal{Cyber Security and Applications}

\begin{document}

\begin{frontmatter}

\title{On the Feasibility of Hybrid Homomorphic Encryption for Intelligent Transportation Systems}

\author[aff1]{Kyle~Yates\tnoteref{tn1,tn2}}
\ead{kjyates@clemson.edu}

\author[aff2]{Abdullah Al~Mamun\corref{cor1}}
\ead{abdullm@clemson.edu/mamun.buet.090@gmail.com}

\author[aff2]{Mashrur~Chowdhury}
\ead{mac@clemson.edu}

\cortext[cor1]{Corresponding author.}

\affiliation[aff1]{organization={School of Mathematical and Statistical Sciences, Clemson University},
            addressline={},
            city={Clemson},
            state={SC},
            postcode={29634},
            country={USA}}

\affiliation[aff2]{organization={Glenn Department of Civil Engineering, Clemson University},
            addressline={},
            city={Clemson},
            state={SC},
            postcode={29634},
            country={USA}}


\begin{abstract}
Many Intelligent Transportation Systems (ITS) applications require strong privacy guarantees for both users and their data. Homomorphic encryption (HE) enables computation directly on encrypted messages and thus offers a compelling approach to privacy-preserving data processing in ITS. However, practical HE schemes incur substantial ciphertext expansion and communication overhead, which limits their suitability for time-critical transportation systems. Hybrid homomorphic encryption (HHE) addresses this challenge by combining a homomorphic encryption scheme with a symmetric cipher, enabling efficient encrypted computation while dramatically reducing communication cost. In this paper, we develop theoretical models of representative ITS applications that integrate HHE to protect sensitive vehicular data. We then perform a parameter-based evaluation of the HHE scheme \textit{Rubato} to estimate ciphertext sizes and communication overhead under realistic ITS workloads. Our results show that HHE achieves orders-of-magnitude reductions in ciphertext size compared with conventional HE while maintaining cryptographic security, making it significantly more practical for latency-constrained ITS communication.
\end{abstract}


\end{frontmatter}

\section{Introduction}
Homomorphic encryption (HE) is an advanced cryptographic tool which allows for computations to be performed directly on encrypted messages (called ciphertexts), ensuring that user information remains secure during computation. Generally speaking, an HE scheme allows for two operations to be performed between ciphertexts: addition and multiplication. HE has several modern and practical applications, especially in secure cloud computing. A user can encrypt their data using HE and outsource computation to the cloud, enabling secure and private data processing. Unfortunately, traditional homomorphic encryption schemes such as BGV \cite{Brakerski2014LeveledFH,cryptoeprint:2011:277}, BFV \cite{cryptoeprint:2012:144}, CKKS \cite{cryptoeprint:2016:421}, and TFHE \cite{cryptoeprint:2018/421} suffer from extreme ciphertext expansion, greatly hindering communication latency and practical bandwidth requirements.

Recently, a new technique known as \textit{hybrid homomorphic encryption} (HHE) has emerged from the HE community. HHE pairs a traditional HE scheme with a symmetric cipher, greatly reducing ciphertext sizes of user messages. Existing HHE schemes include HERA \cite{cryptoeprint:2020/1335}, PASTA \cite{cryptoeprint:2021/731}, and Rubato \cite{cryptoeprint:2022/537}, with a well-written general HHE survey given in \cite{cryptoeprint:2025/071}. HHE offers a promising solution to communication latency issues observed in practical time-sensitive applications.

Many applications which may be worth considering are scenarios arising from Intelligent Transportation Systems (ITS). ITS integrates several advanced electronic communication technologies to improve aspects of transportation such as traffic management and vehicular communications. Several previous works \cite{hannemann2023homomorphicencryptionfeasiblesmart,
10414398,
10498523,
8802665,
icissp21,
9945656,
https://doi.org/10.1002/aisy.202400507,
AMEUR2024151,
9344812,
9523794,
s20154253} study the performance of traditional HE schemes focusing on ITS-specific scenarios. Several other studies \cite{https://doi.org/10.1002/spy2.50,
https://doi.org/10.1049/iet-cps.2019.0042,
MALIK2023500,
9945637,
LIU2025107716,
REN2021105,
10755989} also discuss the Paillier cryptosystem \cite{10.1007/3-540-48910-X_16}. The Paillier cryptosystem is a partially homomorphic cryptosystem, meaning that only addition and scalar multiplication may be performed on ciphertexts. This provides for substantial improvement in latency when compared to traditional HE, but at the cost of allowing for full multiplication operations between ciphertexts. Furthermore, unlike traditional HE the security of the Paillier cryptosystem relies on the hardness of integer factorization and is thus vulnerable to quantum attacks such as Shor's algorithm \cite{shordetailedpaper}. In all the aforementioned works, none discuss the potential use of HHE in ITS applications. HHE may bridge the gap between these two camps of studies, offering a solution which reduces communication latency substantially while at the same time preserving full multiplication operations between ciphertexts.

We should note that a few papers \cite{10.1002/spy2.70109, 10.1145/3770580, 10591464} do discuss hybrid approaches involving homomorphic encryption for ITS applications. The paper \cite{10.1145/3690407.3690562} also develops hybrid approaches with the Paillier cryptosystem for general purposes. However, none of these works use established HHE schemes in developing their models or theory, and instead use the term hybrid approach to indicate the combining HE with other techniques (e.g., federated learning and multi-party computation). This again differs from our approach in this paper, which uses HHE schemes to develop various frameworks for ITS applications and evaluates its theoretical feasibility.

\subsection{Our Contributions}
This paper presents a theoretical study of the feasibility of HHE use for ITS scenarios. After providing the necessary background on HE and HHE, we develop several models of potential ITS applications which use HHE protocols, enhancing user data privacy. Based on parameter choices for Rubato outlined in \cite{cryptoeprint:2022/537}, we theoretically evaluate the expected performance of HHE in these ITS applications based on estimated packet sizes obtained from the Rubato parameters, specifically adhering to bandwidth requirements for ITS scenarios in vehicle-to-everything (V2X) and infrastructure-to-infrastructure (I2I) communication.

\section{Notation and Preliminaries}

\subsection{Notations}
We shall denote $\mathbb{R}$ as the real numbers, $\mathbb{Z}$ as the integers, and $\mathbb{Z}_q = \mathbb{Z}\cap (-q/2,q/2]$ as the ring of integers modulo $q$ centered around $0$. All logarithms are computed in base 2. For $a\in \mathbb{R}$, the notation $\lfloor a \rceil$ denotes the rounding of $a$ to the nearest integer, rounding down in the case of a tie. When $a\in \mathbb{R}^\ell$, we extend the notation $\lfloor a \rceil$ to mean the rounding of each entry. We denote $\mathcal{D}_{\alpha q}$ as the discrete Gaussian distribution over $\mathbb{Z}$ centered at 0. The discrete Gaussian distribution $\mathcal{D}_{\alpha q}$ assigns a probability proportional to $\exp (-\pi a^2/(\alpha q)^2)$ for each $a\in \mathbb{Z}$.

\subsection{Homomorphic Encryption}\label{section.he}
HE is a powerful tool which allows for computations to be performed directly on ciphertexts without any knowledge of private information \cite{Brakerski2014LeveledFH,cryptoeprint:2011:277,cryptoeprint:2012:144,cryptoeprint:2016:421,cryptoeprint:2018/421}. This makes HE a popular and promising solution to several practical challenges. In this paper, we will be especially concerned with theoretical applications of HE in data collection and outsourcing computation.

To better describe the attractive qualities HE has in practical applications, we will outline the basic framework for a scenario in which a user may outsource computation using HE. Let us denote $m_1,\dots,m_{\ell} \in \mathcal{M}$ as a collection of several messages possessed by a user, which are all elements of some message space $\mathcal{M}$. For an arithmetic circuit $C:\mathcal{M}^{\ell}\rightarrow \mathcal{M}$, the user wishes to obtain the value $C(m_1,\dots,m_\ell)$ by outsourcing computation to a server, while attempting not to reveal any of $m_1,\dots,m_{\ell}$ to the server or anyone else. Using a homomorphic encryption algorithm, the user may establish a public key $\texttt{pk}$ and a private key $\texttt{sk}$, then may encrypt each $m_i$ under the public key $\texttt{pk}$ as $\texttt{Enc}_{\texttt{pk}}(m_i)$. The user sends all the $\texttt{Enc}_{\texttt{pk}}(m_i)$'s to the server, where the server may then compute $\texttt{Enc}_{\texttt{pk}}(C(m_1,\dots,m_\ell))$. The server sends the result $\texttt{Enc}_{\texttt{pk}}(C(m_1,\dots,m_\ell))$ back to the user, which the user can decrypt to obtain $C(m_1,\dots,m_\ell)$ using the secret key $\texttt{sk}$. In this model, the user obtain their desired result without the server ever accessing the secret key $\texttt{sk}$ or any plaintext corresponding to a ciphertext $\texttt{Enc}_{\texttt{pk}}(m_i)$.

An HE scheme which we will take particular note of is the CKKS scheme \cite{cryptoeprint:2016:421}. The CKKS scheme allows for homomorphic computation (i.e., computation on ciphertexts) on real numbers with approximate floating-point arithmetic, and is currently the only scheme to do so. Other widely used schemes such as BGV \cite{Brakerski2014LeveledFH,cryptoeprint:2011:277}, BFV \cite{cryptoeprint:2012:144}, and TFHE \cite{cryptoeprint:2018/421} only allow exact arithmetic over $\mathbb{Z}_q$ or $\mathbb{Z}$ for chosen parameter $q$. This makes CKKS an immensely popular HE scheme, especially for applications which require floating-point arithmetic such as privacy-preserving machine learning.

The theoretical security of all modern HE schemes is based on the hardness of learning with errors (LWE). For randomly chosen $s\in \mathbb{Z}_q^n$, the LWE problem is to recover $s$ given many pairs of the form $(a,a\cdot s + e \mod q)$ where $a$ is selected uniform randomly from $\mathbb{Z}_q^n$ and $e$ is sampled from $\mathcal{D}_{\alpha q}$. An additional advantage of HE is that since the underlying security is based on the hardness of LWE, HE is presumed to be secure against attacks by quantum computers. Thus, HE falls under the broader umbrella of post-quantum cryptography, which describes cryptographic protocols which can be implemented classically but are resistant to quantum attacks based on algorithms such as Shor's algorithm \cite{shordetailedpaper} and Grover's algorithm \cite{groverexpandedpaper}.

\subsection{Hybrid Homomorphic Encryption}
A major disadvantage of traditional HE is that these schemes generally suffer from tremendous ciphertext expansion and overhead, rendering them limited for practical application. Said otherwise, for our described scenario involving a user and server in Section \ref{section.he}, the information which the user and server send back and forth is enormous compared to traditional encryption schemes used in practice such as RSA and AES.

To address the issue of huge ciphertext sizes, the topic of HHE has emerged in recent years. HHE takes a more complicated approach than HE, but with the same goal in mind of outsourcing computation to the server. This is done by pairing an HE algorithm with a symmetric encryption algorithm, drastically reducing the sizes of ciphertexts to be sent from the user to the server. To better describe the general HHE framework and differences between HE and HHE, let us revise the notation from the previous section and introduce some new notation.

Suppose we now have two encryption algorithms: an HE algorithm denoted $\texttt{Enc}^{\texttt{Hom}}$ and a symmetric encryption algorithm denoted $\texttt{Enc}^{\texttt{Sym}}$. The HE algorithm has an associated public key $\texttt{pk}$ and private key $\texttt{sk}$, while the symmetric encryption algorithm has a single (private) key $\texttt{k}$. We denote $\texttt{Enc}_{\texttt{pk}}^{\texttt{Hom}}(v)$ as the encryption of $v$ using the homomorphic encryption algorithm under public key $\texttt{pk}$, and $\texttt{Enc}_{\texttt{k}}^{\texttt{Sym}}(v)$ as the encryption of $v$ using the symmetric encryption algorithm under symmetric key $\texttt{k}$. For a circuit $C$, we shall denote $\texttt{Eval}^{\texttt{Hom}}(C)$ as the homomorphic evaluation of the circuit. Lastly, we shall denote $\texttt{Dec}^{\texttt{Sym}}$ as the symmetric decryption circuit (i.e., if $\texttt{ct} = \texttt{Enc}_{\texttt{k}}^{\texttt{Sym}}(v)$, then $v = \texttt{Dec}_{\texttt{k}}^{\texttt{Sym}}(\texttt{ct})$).

Our general goal with HHE will be the same as before with HE: we have a user who wishes to export computation to a server, while maintaining privacy of their messages and their private key. For a collection $m_1,\dots,m_{\ell}$ of user messages, this is still done by the server first obtaining all the $\texttt{Enc}^{\texttt{Hom}}_{\texttt{pk}}(m_i)$'s then computing $\texttt{Enc}^{\texttt{Hom}}_{\texttt{pk}}(C(m_1,\dots,m_\ell))$. However, the main difference with HHE is now how the server obtains all the $\texttt{Enc}^{\texttt{Hom}}_{\texttt{pk}}(m_i)$'s. We describe the procedure in Protocol \ref{hhe.protocol} for a single message $m$, which describes a two-phase protocol involving the user and the server. Phase 1 belongs to the user, while phase 2 belongs to the server.

\renewcommand{\figurename}{Protocol}
\begin{figure}[t]
\centering
\renewcommand{\arraystretch}{1.15}
\begin{tabular}{|p{2.0cm}p{10.2cm}|}
\hline
\multicolumn{2}{|l|}{\textbf{Phase 1: User}}\\
\hline
Step 1. & Encrypt $m$ using the symmetric encryption algorithm under symmetric key $\texttt{k}$ to obtain $\texttt{Enc}_{\texttt{k}}^{\texttt{Sym}}(m)$.\\
Step 2. & Encrypt $\texttt{k}$ using the homomorphic encryption algorithm under public key $\texttt{pk}$ to obtain $\texttt{Enc}_{\texttt{pk}}^{\texttt{Hom}}(\texttt{k})$.\\
Step 3. & Send $\texttt{Enc}_{\texttt{k}}^{\texttt{Sym}}(m)$ and $\texttt{Enc}_{\texttt{pk}}^{\texttt{Hom}}(\texttt{k})$ to the server.\\
\hline
\multicolumn{2}{|l|}{\textbf{Phase 2: Server}}\\
\hline
Step 1. & Encrypt $\texttt{Enc}_{\texttt{k}}^{\texttt{Sym}}(m)$ using the homomorphic encryption algorithm to obtain $\texttt{Enc}_{\texttt{pk}}^{\texttt{Hom}}(\texttt{Enc}_{\texttt{k}}^{\texttt{Sym}}(m))$.\\
Step 2. & Homomorphically evaluate the symmetric decryption circuit, computing $\texttt{Eval}^{\texttt{Hom}}(\texttt{Dec}^{\texttt{Sym}})$ with inputs $\texttt{Enc}_{\texttt{pk}}^{\texttt{Hom}}(\texttt{Enc}_{\texttt{k}}^{\texttt{Sym}}(m))$ and $\texttt{Enc}_{\texttt{pk}}^{\texttt{Hom}}(\texttt{k})$ to obtain $\texttt{Enc}_{\texttt{pk}}^{\texttt{Hom}}(m)$.\\
\hline
\end{tabular}
\renewcommand{\arraystretch}{1}
\caption{Hybrid homomorphic encryption protocol.}
\label{hhe.protocol}
\end{figure}

Using this model, observe that the server still obtains all the ciphertexts $\texttt{Enc}_{\texttt{pk}}^{\texttt{Hom}}(m_i)$ for each message $m_i$. However, the user only transmits a symmetric encryption $\texttt{Enc}_{\texttt{k}}^{\texttt{Sym}}(m_i)$ of each $m_i$, and one homomorphic encryption $\texttt{Enc}_{\texttt{pk}}^{\texttt{Hom}}(\texttt{k})$ of the symmetric key. This significantly decreases the total amount of data being transmitted when several messages are encrypted and sent since symmetric key algorithms have much smaller ciphertexts than HE algorithms. Heavy computation is also diverted to the server when possible, such as the computation in Steps 1 and 2 of Phase 2 in Protocol \ref{hhe.protocol}.

For purposes of practical applications, it will be useful for us to consider the purpose of Protocol \ref{hhe.protocol} above in regards to improving communication latency. We therefore point the reader to Figure \ref{hhe.figure}, which visually depicts Protocol \ref{hhe.protocol} and the corresponding communications. Here, any algorithms are denoted by a solid box around the border, whereas data is not. From this figure, we get a clearer sense of the communication latency we may observe. The user must send $\texttt{Enc}_{\texttt{pk}}^{\texttt{Hom}}(\texttt{k})$ and $\texttt{Enc}_{\texttt{k}}^{\texttt{Sym}}(m)$ (or usually, many $\texttt{Enc}_{\texttt{k}}^{\texttt{Sym}}(m_i)$'s) across the communication channel to the server. The notable takeaway here is that generally each $\texttt{Enc}_{\texttt{k}}^{\texttt{Sym}}(m)$ is much smaller than $\texttt{Enc}_{\texttt{pk}}^{\texttt{Hom}}(m)$ would be, drastically improving communication latency compared to a standard HE model.

\setcounter{figure}{0}
\renewcommand{\figurename}{Figure}
\begin{figure}[t]
\centering
\scalebox{0.8}{
\begin{tikzpicture}
\draw (-2,-2.5) node(a) {};
\draw (-2,5) node(b) {};
\draw[dashed,thick] (a.north) -- (b.south);
\draw (2,-2.5) node(c) {};
\draw (2,5) node(d) {};
\draw[dashed,thick] (c.north) -- (d.south);
\draw (0,5.5) node(com1) {\textsf{Communication}};
\draw (0,5) node(com2) {\underline{\textsf{Channel}}};

\draw (-5,5.25) node(user) {\underline{\textsf{User}}};
\draw (5,5.25) node(server) {\underline{\textsf{Server}}};

\draw (-6.5,4.5) node(1) {$m$};
\draw (-3.5,4.5) node(2) {$\texttt{k}$};

\draw (-8.0,2.5) node(3a) {${\texttt{k}}$};
\draw (-6.5,2.5) node(3) [draw] {$\texttt{Enc}^{\texttt{Sym}}$};
\draw (-5.0,2.5) node(4a) {${\texttt{pk}}$};
\draw (-3.5,2.5) node(4) [draw] {$\texttt{Enc}^{\texttt{Hom}}$};

\draw[->] (1.south) -- (3.north);
\draw[->] (2.south) -- (4.north);
\draw[->] (3a.east) -- (3.west);
\draw[->] (4a.east) -- (4.west);

\draw (-6.5,0.5) node(5) {$\texttt{Enc}_{\texttt{k}}^{\texttt{Sym}}(m)$};
\draw (-3.5,0.5) node(6) {$\texttt{Enc}_{\texttt{pk}}^{\texttt{Hom}}(\texttt{k})$};

\draw[->] (3.south) -- (5.north);
\draw[->] (4.south) -- (6.north);

\draw (3.5,-0.5) node(7a) {$\texttt{pk}$};
\draw (3.5,-1.5) node(7) [draw] {$\texttt{Enc}^{\texttt{Hom}}$};

\draw[->] (7a.south) -| (7.north);
\draw[->] (5.south) |- (7.west);

\draw (6.5,-1.5) node(9) {$\texttt{Enc}_{\texttt{pk}}^{\texttt{Hom}}(\texttt{Enc}_{\texttt{k}}^{\texttt{Sym}}(m))$};
\draw (6.5,0.5) node(10) [draw] {$\texttt{Eval}^{\texttt{Hom}}(\texttt{Dec}^{\texttt{Sym}})$};

\draw[->] (7.east) |- (9.west);
\draw[->] (6.east) |- (10.west);
\draw[->] (9.north) |- (10.south);

\draw (6.5,4.5) node(11) {$\texttt{Enc}_{\texttt{pk}}^{\texttt{Hom}}(m)$};

\draw[->] (10.north) |- (11.south);
\end{tikzpicture}
}
\caption{Hybrid homomorphic encryption model.}
\label{hhe.figure}
\end{figure}
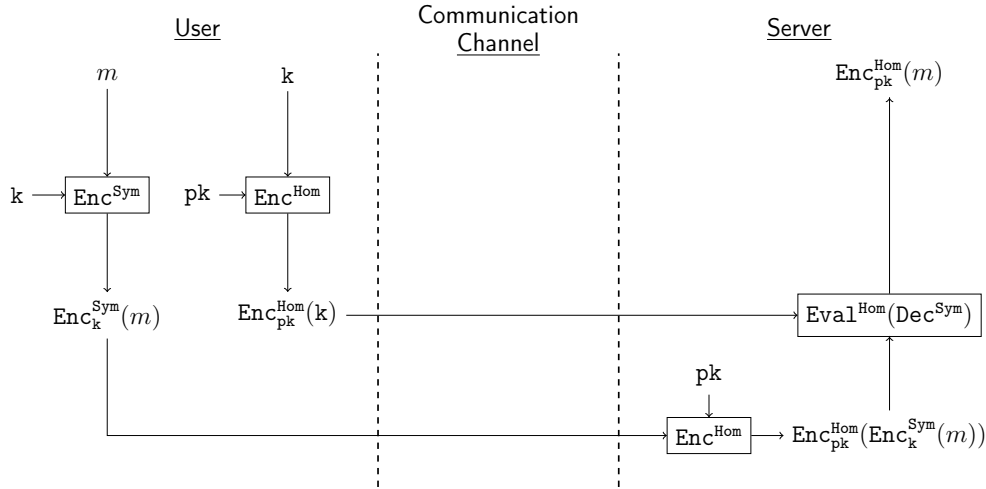

Unfortunately, there are a few downsides with using HHE. First, the symmetric cipher used must be one for which $\texttt{Eval}^{\texttt{Hom}}(\texttt{Dec}^{\texttt{Sym}})$ may be efficiently computed. Standard and commonly used symmetric ciphers often can not have their decryption circuits homomorphically evaluated efficiently (see, e.g., the study on AES \cite{10.1007/978-3-642-32009-5_49}). Thus, much of the ongoing research in the literature is devoted to developing symmetric ciphers whose decryption circuits may be homomorphically evaluated efficiently.

We should remark that although this describes the overall ideas behind HE and HHE, several details have been omitted. In particular, an additional key known as the \textit{evaluation key} must be published by the user in order for the server to evaluate any circuits homomorphically. We refer the reader to \cite{cryptoeprint:2025/071} for more details on the evaluation key and other aspects of HHE which we omitted in this study.

\subsubsection{Rubato}\label{section.rubato}
Introduced in 2022, the Rubato scheme \cite{cryptoeprint:2022/537} is an HHE scheme allowing for approximate homomorphic computation, which utilizes aspects of both CKKS and BFV. Rubato builds on a Real-to-Finite-field (\textsf{RtF}) transciphering framework introduced in HERA \cite{cryptoeprint:2020/1335}. The \textsf{RtF} framework in Rubato uses a symmetric cipher $\textsf{E}$ equipped with key $\texttt{k}$ and additional Gaussian noise to form the symmetric encryptions $\texttt{Enc}_{\texttt{k}}^{\texttt{Sym}}(m)$. Notice here that we draw a distinction between $\textsf{E}$ and $\texttt{Enc}_{\texttt{k}}^{\texttt{Sym}}(m)$. The cipher $\textsf{E}$ is used as a building block for our symmetric encryptions, as we will explain shortly. It should be noted though that the key $\texttt{k}$ of $\textsf{E}$ is the only symmetric key we use.

Although we omit most details of Rubato and refer the reader to \cite{cryptoeprint:2022/537} for a more comprehensive understanding of the scheme, let us describe the main idea of the Rubato design. We denote $\textsf{E}$ as the Rubato cipher. The Rubato cipher inputs a nonce $\texttt{nc}\in \{ 0,1\}^{\lambda}$ and a symmetric key $\texttt{k}\in \mathbb{Z}_q^n$ for security parameter $\lambda$ and prime $q$. The output of the cipher is a keystream $z$, which we shall write as $z=\textsf{E}_\texttt{k}(\texttt{nc})$. The construction of $\textsf{E}$ is out of the scope of this paper, and thus we refer the reader to \cite{cryptoeprint:2022/537} for details on the construction of $\textsf{E}$. Rubato encryption then takes the form of a ciphertext $\texttt{ct} = \lfloor \Delta \cdot m \rceil + z \mod q$, where $m\in \mathbb{R}^{\ell}$ is our message and $\Delta > 0$ is a positive scaling factor. Rubato and HERA are quite similar in their encryption style. However, we should note that in Rubato the keystream $z$ contains Gaussian noise. That is, the cipher $\textsf{E}$ is a combination of a more conventional symmetric cipher and extra Gaussian noise. This contributes to a lower number of multiplications in the Rubato cipher needed for encryption and (homomorphic) decryption, improving practical performance when compared to other HE and HHE schemes. However, this Gaussian noise also contributes to additional precision loss, making the Rubato cipher inappropriate for situations which require exact transciphering.

\section{Applications of HHE in ITS}\label{section.ITSapplications}
Prior work examining the feasibility of homomorphic encryption in vehicular networks has shown that fully homomorphic encryption schemes (FHE) introduce substantial communication and computational overhead, making them impractical for latency-sensitive broadcast channels, such as vehicle-to-vehicle (V2V) and vehicle-to-infrastructure (V2I) exchanges. As summarized by Sun et al., the ciphertext expansion and processing demands of FHE conflict directly with the tight timing constraints that govern real-time vehicular communication \cite{s20154253}. More recent experimental studies on post-quantum HE in transportation environments reinforce this observation, demonstrating that lattice-based HE ciphertexts can reach into hundreds of kilobytes and consequently cause packet fragmentation, increased queuing, and significant delays (3--31\,s depending on the HE scheme and the number of HE operations) over both wireless and wired links \cite{mamun2025experimentalevaluationpostquantumhomomorphic}. This overhead conflicts with the 10\,Hz Basic Safety Message (BSM) transmission interval required by SAE J2735 and J2945/1, where slight latency increases can affect the reliability of safety functions \cite{SAE,SAE2}. For this reason, existing uses of HE in ITS have focused on infrastructure-to-infrastructure (I2I) data flows, where road-side units (RSUs) periodically forward aggregated BSM data to cloud or analytics platforms and where the timing requirements are less stringent than those governing vehicle-level safety applications \cite{icissp21}.

HHE offers a promising alternative for I2I communication by replacing large homomorphic ciphertexts with compact symmetric ciphertexts, thereby eliminating much of the communication overhead that limits pure HE \cite{cryptoeprint:2025/071}. Offloading the homomorphic workload to the cloud further reduces uplink bandwidth demands and minimizes fragmentation. Together, these features suggest that HHE can achieve lower end-to-end latency while maintaining privacy, improving its suitability for I2I links and for understanding whether such approaches could eventually extend to more time-critical V2X communication links, such as V2V and V2I.

Within an ITS architecture, HHE integrates naturally with the RSU--Cloud--TMC communication pipeline, where TMC denotes the Traffic Management Center. RSUs collect and authenticate BSMs from multiple vehicles using Security Credential Management System (SCMS) credentials and aggregate these BSMs, including location, speed, and acceleration information, after verifying their authenticity \cite{noauthor_scms_2022}. Because HHE is applied only within the RSU backhaul path (after BSM collection), the mandated plaintext broadcast and timing requirements of V2X safety messages remain fully preserved. In this workflow, the RSU acts as the data provider, the cloud functions as the untrusted evaluator, and the TMC typically holds the homomorphic secret key and decrypts the final analytical result. Although the TMC is the default decryptor, the architecture can also accommodate RSU-side decryption when localized decisions or edge responsiveness are required.

\begin{figure}[t]
\centering
\includegraphics[width=\textwidth]{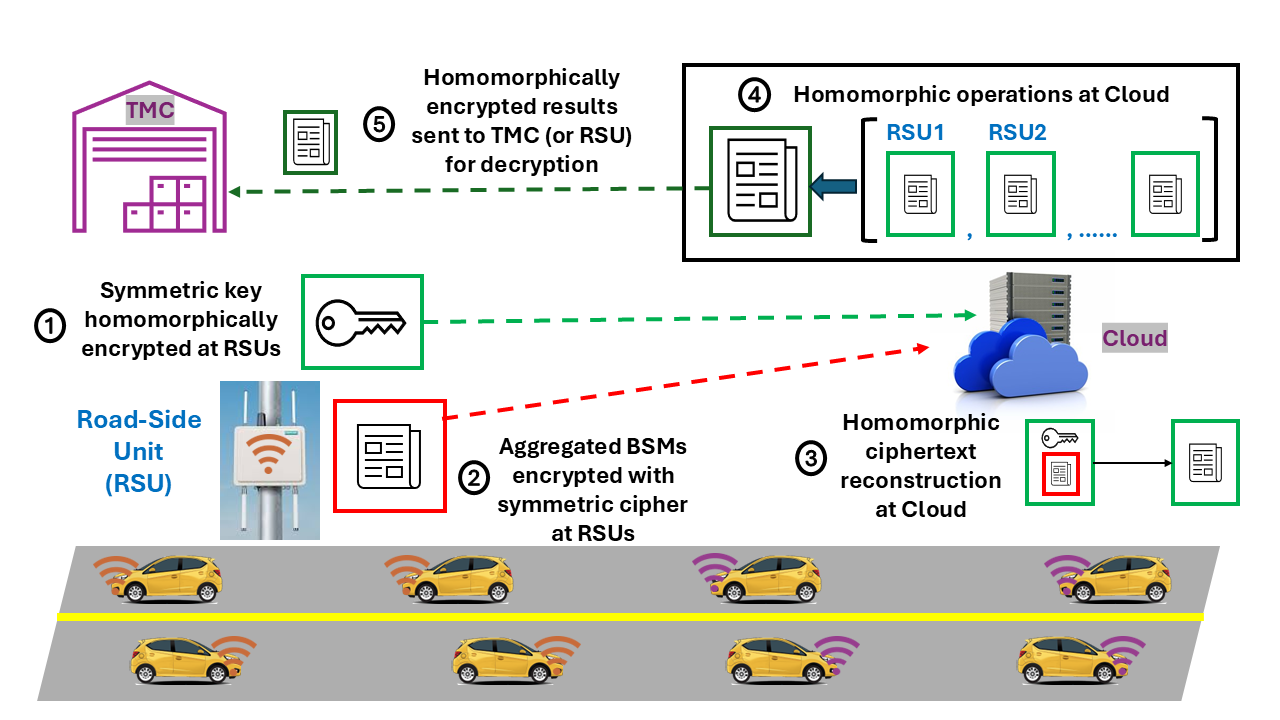}
\caption{Overview of the HHE-enabled I2I workflow. Each RSU first homomorphically encrypts its symmetric key (Step 1). During operation, aggregated BSMs are encrypted using lightweight symmetric cryptography and uploaded to the cloud (Step 2). The cloud reconstructs the homomorphic ciphertext (Step 3) and performs encrypted analytics without decryption (Step 4). The encrypted results are sent back to the TMC or RSU for decryption (Step 5).}
\label{hhe.its.figure}
\end{figure}

A representative HHE-enabled I2I workflow is illustrated in Figure \ref{hhe.its.figure}. The process begins with an offline initialization phase in which each RSU generates a symmetric key and sends its homomorphically encrypted form to the cloud. During regular operation, multiple RSUs encrypt their aggregated BSMs using a lightweight symmetric cipher which is compatible for HHE (e.g., Rubato) and transmit these compact ciphertexts to the cloud. Because these uploads consist of typically at most 100--200 bytes, the RSU--Cloud backhaul fits within the bandwidth limits of typical Wi-Fi or Ethernet deployments. This uplink transmission is one of the two communication points that meaningfully contributes to end-to-end latency. Importantly, these symmetric ciphertexts are orders of magnitude smaller than the multi-hundred-kilobyte ciphertexts produced by traditional post-quantum HE schemes, and this size reduction is the key reason HHE should theoretically achieve significantly better communication performance.

After receiving the encrypted data, the cloud homomorphically decrypts the symmetric ciphertexts using the HE-encrypted keys supplied during initialization, thereby reconstructing homomorphic ciphertexts without learning the underlying data. Using these ciphertexts, the cloud carries out encrypted analytics, such as computing average speed, queue length, or corridor-level congestion indices, through homomorphic additions and limited multiplications. Once the computation is complete, the cloud produces a single aggregated homomorphic ciphertext and returns it to the key holder, typically the TMC though RSU-side decryption remains an option. This downstream transmission of a single large ciphertext represents the second key communication step influencing end-to-end latency. However, unlike pure HE workflows, only one such ciphertext is transmitted per computation cycle rather than one per RSU or per measurement. The TMC/RSU decrypts this final ciphertext and distributes the analytic result for operational decision-making.

Section 4 will build on this framework by discussing and evaluating parameter choices for HHE, particularly using the Rubato scheme. We will outline the corresponding expected ciphertext sizes from various parameter choices and use this as the main metric for determining the feasibility of HHE for ITS application scenarios such as that outlined in Figure \ref{hhe.its.figure}.

\section{Theoretical Evaluation of HHE for ITS}
In this section, we will provide a theoretical evaluation of the feasibility of HHE for use in ITS. We specifically focus on the HHE scheme Rubato, as it offers a nice balance between small parameter sizes and fast homomorphic evaluation of the symmetric decryption circuit. Recall that in our overview of Rubato in Section \ref{section.rubato}, a distinguishing factor between Rubato and HERA was the additional presence of Gaussian noise used to hide messages in the Rubato cipher $\textsf{E}$. Let us assume this Gaussian noise in the Rubato cipher is sampled specifically from $\mathcal{D}_{\alpha q}$ for parameter $\alpha q$. We then outline the small parameter choices for Rubato in Table \ref{rubato.params} we will use, which is a simplified version of Table 2 in \cite{cryptoeprint:2022/537}. Here, note that $\lambda$ is the security level and the parameters $n,\ell,q$ and $\alpha q$ are the same as in the Rubato explanation we provided during Section \ref{section.rubato}.

\begin{table}[h!]
\centering
\renewcommand{\arraystretch}{1.15}
\caption{Rubato parameters (adapted from \cite{cryptoeprint:2022/537}).}
\vspace{4pt}
\begin{tabular}{|c|c|c|c|c|c|}
\hline
Rubato Parameters & $\lambda$ & $n$ & $\ell$ & $\lceil \log q \rceil$ & $\alpha q$ \\
\hline
\texttt{Par-80S}   & $80$  & $16$ & $12$ & $26$ & $11.1$ \\
\texttt{Par-80M}   & $80$  & $36$ & $32$ & $25$ & $2.7$ \\
\texttt{Par-80L}   & $80$  & $64$ & $60$ & $25$ & $1.6$ \\
\texttt{Par-128S}  & $128$ & $16$ & $12$ & $26$ & $10.5$ \\
\texttt{Par-128M}  & $128$ & $36$ & $32$ & $25$ & $4.1$ \\
\texttt{Par-128L}  & $128$ & $64$ & $60$ & $25$ & $4.1$ \\
\hline
\end{tabular}
\label{rubato.params}
\end{table}

An analysis on latency and ciphertext sizes for various parameters for the $\textsf{RtF}$ framework, including those in Table~\ref{rubato.params}, is given in Tables~5--7 of \cite{cryptoeprint:2022/537}. We will perform some of our own theoretical analysis using these parameters, with a specific focus on applications for ITS as outlined in Section~\ref{section.ITSapplications}. For the parameters outlined in Table~\ref{rubato.params}, we will analyze our expected ciphertext sizes and effective plaintext payloads.

Based on the parameters given in Table~\ref{rubato.params}, it is straightforward to estimate the maximum size of a ciphertext. For example, let us suppose we have parameters $n=16$, $\ell=12$, and $\lceil \log q \rceil = 26$ (i.e., \texttt{Par-80S} or \texttt{Par-128S}). Recall that a Rubato encryption $\texttt{ct}\in\mathbb{Z}_q^{\ell}$ satisfies
\[
\texttt{ct} = \lfloor \Delta \cdot m \rceil + z \pmod q,
\]
where $m\in \mathbb{R}^{\ell}$ is a message, $\Delta$ is a positive scaling parameter, and $q$ is prime. Each coefficient of $\texttt{ct}$ therefore lies in $(-q/2,q/2]$, and for $\lceil \log q \rceil = 26$ each coefficient is represented by at most 27 bits. With $\ell=12$ coefficients, each ciphertext is represented by at most $12\times27=324$ bits, or 41~bytes. The corresponding Rubato ciphertext sizes for all parameter sets are summarized in Table~\ref{rubato.sizes}.

Since Rubato encrypts $\ell$ packed plaintext slots per ciphertext \cite{cryptoeprint:2022/537}, the effective data-carrying capacity depends on how each slot is quantized. If each plaintext slot is represented using $B$-bit fixed-point encoding, then the maximum plaintext payload per Rubato ciphertext is $\lceil \ell B / 8 \rceil$ bytes. In ITS applications, quantities such as speed, flow, and density can be accurately represented using 16-bit fixed-point precision, and we therefore adopt $B=16$ bits per slot when computing plaintext payloads in Table~\ref{rubato.sizes}. Further precaution should be taken when fine-tuning parameters for practical application. One should ensure $\Delta$ is chosen appropriately according to the guidelines in \cite{cryptoeprint:2022/537}. More specifically, $\Delta$ should be chosen as $\Delta = q/(16\cdot b)$ for $b$ satisfying $\vert|{m}\vert|_1\leq b$, where $\vert|{m}\vert|_1$ denotes the standard 1-norm.

\begin{table}[h!]
\centering
\renewcommand{\arraystretch}{1.15}
\caption{Rubato ciphertext sizes and effective plaintext payload for ITS telemetry (assuming $B=16$-bit fixed-point encoding per slot).}
\vspace{4pt}
\begin{tabular}{|c|c|c|c|}
\hline
Rubato Parameters & $\ell$ & Payload (bytes) & Rubato $\texttt{ct}$ size (bytes) \\
\hline
\texttt{Par-80S}   & 12 & 24  & 41  \\
\texttt{Par-80M}   & 32 & 64  & 104 \\
\texttt{Par-80L}   & 60 & 120 & 195 \\
\texttt{Par-128S}  & 12 & 24  & 41  \\
\texttt{Par-128M}  & 32 & 64  & 104 \\
\texttt{Par-128L}  & 60 & 120 & 195 \\
\hline
\end{tabular}
\label{rubato.sizes}
\end{table}

We should note that the ciphertext sizes in Table \ref{rubato.sizes} only account for the ciphertexts representing user messages. The sizes do not reflect the entire information flow that must be transmitted in order to successfully use HHE. Particularly, if each user encrypts $r$ messages for sending, each user must also generate and send $r$ nonces to the server totaling an additional $r\cdot \lambda$ transmitted bits for a $\lambda$-bit security level. Each user must also, after generating their symmetric key $\texttt{k}$, send the server $\texttt{Enc}_{\texttt{pk}}^{\texttt{Hom}}(\texttt{k})$. The size of $\texttt{Enc}_{\texttt{pk}}^{\texttt{Hom}}(\texttt{k})$ can be huge, greatly hindering the advantages of HHE if done naively.

For our ITS applications, this extra exchanging of information does not pose practical issues. We shall explain this using the scenario outlined in Figure \ref{hhe.its.figure}. Recall that in Step 1, the RSUs encrypt the symmetric keys using a homomorphic encryption algorithm and send to the cloud. These encrypted keys, along with the nonces, can be generated and sent in an offline phase before any other transmissions or computations occur. Hence, communication overhead is not an issue for this step. Step 2 is the primary concern for time-sensitive transmissions of vehicle telemetry between RSUs and the cloud server. The data payload sizes transmitted in this step are given by the ciphertext sizes in Table \ref{rubato.sizes}. These sizes are much smaller than traditional HE, highlighting the main advantage of HHE for use in ITS.

A comparison of ciphertext expansion further illustrates why HHE is particularly well suited for ITS communication. Traditional HE schemes produce very large ciphertexts. For example, a single BFV ciphertext is 131{,}939 bytes, BGV reaches 394{,}573 bytes, and CKKS ciphertexts range from 787{,}791 bytes (add-only) up to more than 1{,}050{,}129 bytes when multiplications are included under 128-bit post-quantum security \cite{cryptoeprint:2024/463}. In the experimental setting of \cite{mamun2025experimentalevaluationpostquantumhomomorphic}, where each plaintext ITS message was 200~bytes, which is representative of the payload size of a BSM defined under the SAE~J2735 framework \cite{2011-01-0584}, this corresponds to ciphertext expansion factors of approximately 660 for BFV, 1{,}970 for BGV, and 3{,}940--5{,}250 for CKKS. In contrast, Rubato, the representative HHE scheme considered in this study, produces ciphertexts of only 41--195 bytes while carrying 24--120 bytes of ITS telemetry, corresponding to an expansion factor of only about 1.6--1.7. These quantitative differences are summarized in Table~\ref{mtu.expansion}.

\begin{table}[h!]
\centering
\small
\renewcommand{\arraystretch}{1.15}
\setlength{\tabcolsep}{4pt}
\caption{Ciphertext expansion and MTU fragmentation for pure HE versus HHE (Rubato). BFV, BGV, and CKKS values correspond to 200-byte ITS plaintexts from \cite{mamun2025experimentalevaluationpostquantumhomomorphic}. Rubato payloads assume $B=16$-bit fixed-point encoding per slot. Fragments are based on 1400-byte MTU. add+mul: addition+multiplication.}
\vspace{4pt}
\begin{tabular}{|l|r|r|r|r|}
\hline
Scheme & \multicolumn{1}{c|}{Plaintext} & \multicolumn{1}{c|}{Ciphertext} & \multicolumn{1}{c|}{Expansion} & \multicolumn{1}{c|}{MTU Fragments} \\
       & \multicolumn{1}{c|}{(B)}       & \multicolumn{1}{c|}{(B)}        & \multicolumn{1}{c|}{(times)}   &                                      \\
\hline
BFV    & 200 & 131939  & 660  & 95 \\
BGV    & 200 & 394573  & 1973 & 284 \\
CKKS (add) & 200 & 787791  & 3939 & 566 \\
CKKS (add+mul) & 200 & 1050129 & 5251 & 754 \\
\hline
Rubato (Par-80S) & 24  & 41  & 1.7 & 1 \\
Rubato (Par-80M) & 64  & 104 & 1.6 & 1 \\
Rubato (Par-80L) & 120 & 195 & 1.6 & 1 \\
\hline
\end{tabular}
\label{mtu.expansion}
\end{table}

On standard Ethernet and Wi-Fi links, the maximum transmission unit (MTU) is approximately 1{,}500 bytes, so any packet larger than this limit must be fragmented at the IP layer, incurring additional headers, buffering, retransmissions, and reassembly delay. As shown in Table~\ref{mtu.expansion}, pure HE ciphertexts such as BFV, BGV, and CKKS therefore require between roughly 90 and more than 700 MTU-sized fragments per encrypted message, whereas Rubato ciphertexts always fit within a single MTU-bounded packet. This theoretical fragmentation burden is consistent with the measurements of Mamun \emph{et al.}, who observed multi-second round-trip delays for BFV and CKKS due to extensive UDP fragmentation \cite{mamun2025experimentalevaluationpostquantumhomomorphic}, while HHE-scale ciphertexts would avoid these delays by eliminating fragmentation entirely on the RSU--cloud backhaul.

This improvement in communication latency is especially important given the communication patterns of ITS systems. RSUs receive BSMs at 10\,Hz (10 messages per second) and must forward aggregated information to the cloud frequently enough to avoid data loss and preserve the responsiveness of real-time and near-real-time applications. Uplink transmission therefore occurs continuously, while the return path from the cloud to the RSU or TMC can occur less frequently and still satisfy operational timelines. In such a setting, the compact ciphertexts produced by HHE significantly reduce uplink burden, making high-frequency encrypted data uploads feasible without overwhelming the backhaul network.

We should mention that, although HHE provides significant advantage over traditional HE for ITS, it is not a perfect solution. Due to the extra use of a symmetric cipher, an additional homomorphic evaluation of the symmetric circuit must occur in Step 3 of Figure \ref{hhe.its.figure}. This is a necessary tradeoff for the improved communication costs achieved in Step 2. Furthermore, a homomorphically encrypted ciphertext must still be transmitted between the cloud and TMC/RSU in Step 5, slowing the overall workflow due to the large ciphertext expansion that HE ciphertexts endure.

\section{Conclusion and Future Work}
In this paper, we have performed a theoretical evaluation of HHE for application in ITS. We develop models that leverage HHE protocols tailored for ITS applications requiring data privacy, providing enhancements over current systems in which data remain exposed to adversaries or must be decrypted and accessed by cloud servers to perform analytics. Our findings point to the very real potential HHE has in ITS, indicating possible drastic improvement in communication latency in practical scenarios due to the much smaller ciphertext size transmitted between RSUs and the cloud. Although our study points to this outcome theoretically, an experimental study using simulations and/or hardware remains to be conducted. We leave this for future work, hoping that experimental data will supplement evidence of HHE being used practically in ITS.

\section*{Acknowledgements}

This research was supported by the National Center for Transportation
Cybersecurity and Resiliency (TraCR), a U.S.\ Department of Transportation National
University Transportation Center, headquartered at Clemson University, Clemson,
South Carolina, USA. Any opinions, findings, conclusions, and recommendations
expressed in this material are those of the authors and do not necessarily reflect
the views of TraCR. The U.S.\ Government assumes no liability for the contents or
use thereof.

\section*{Funding}

This work was funded by the National Center for Transportation Cybersecurity
and Resiliency (TraCR).

\bibliographystyle{elsarticle-num}
\bibliography{bibliography}

\end{document}